\documentclass{iopart}

\usepackage[dvips]{graphicx,color}
\newcommand{\NF}{\textit{Nucl. Fusion} }
\newcommand{\PP}{\textit{Phys. Plasmas} }
\newcommand{\PFB}{\textit{Phys. Fluids B} }

\begin{document}

\title{The plasma boundary in Single Helical Axis RFP plasmas}
\author{E.~Martines, R.~Lorenzini, B.~Momo, 
        S.~Munaretto, P.~Innocente, M.~Spolaore}
\address{Consorzio RFX, Associazione Euratom-ENEA sulla Fusione, 
         corso Stati Uniti 4, 35127 Padova, Italy}
\ead{emilio.martines@igi.cnr.it}

\begin{abstract}
Single Helical Axis (SHAx) states obtained in high current reversed field pinch (RFP) plasmas display, aside from a dominant mode in the m=1 spectrum, also a dominant m=0 mode, with the same toroidal mode number as the m=1 one. The two modes have a fixed phase relationship. The island chain created by the m=0 mode across the reversal surface gives rise, at shallow reversal of the toroidal field, to an X-point structure which separates the last closed flux surface from the first wall, creating a divertor-like configuration. The plasma-wall interaction is found to be related to the connection length of the field lines intercepting the wall, which displays a pattern modulated by the dominant mode toroidal periodicity. This configuration, which occurs only for shallow toroidal field reversal, could be exploited to realize an island divertor in analogy to stellarators. 
\end{abstract}


\section{Introduction}
A proper control of the way in which the plasma interacts with the vacuum chamber wall and the plasma-facing components is a primary issue for high-temperature magnetically confined plasmas of relevance to nuclear fusion research \cite{stangeby}. 
Indeed, a magnetically confined plasma is an open system where particles and energy are continuously injected from outside and the non-perfect confinement properties of the plasma give rise to fluxes of these quantities coming out of the plasma boundary. 
The position and shape of the plasma boundary, identified by the Last Closed Flux Surface (LCFS), that is the outermost magnetic surface not intersecting any solid object, plays a crucial role in determining how the outcoming fluxes interact with the plasma-facing components. 

Generally speaking, two different topologies of the LCFS can be identified. In the simplest configuration the LCFS touches a solid object, called limiter, which defines its position. This direct contact leads to intense heat loads and a high degree of particle recycling. The second, more sophisticated approach consists in creating a divertor, that is a magnetic structure with one or more X-points, which act as links between a LCFS not touching any solid object and open field lines which are diverted towards appropriate plates, on which heat and particle fluxes impinge. These plates can be located far away from the LCFS, so that neutrals resulting from plasma neutralization and sputtering processes can be ionized before entering the confined plasma, thus resulting in a low recycling of the main gas and in a lower impurity content of the plasma.

In the tokamak \cite{wesson}, which is the most advanced toroidal configuration and on which the ITER project is based, the divertor concept has been adopted since long time, and is an integral part of any modern machine design. The layout which has proved to be the most effective is that of a single null poloidal divertor. In the stellarator the problem of designing a divertor configuration has been undertaken more recently \cite{review_divertor_stellarator}. Being it an inherently 3D configuration, a more complex structure has to be adopted, and work is still in progress for defining the best solution. The concepts presently under consideration, the island divertor and the helical divertor, are both based on forming low-order magnetic island chains in the outer plasma region, and exploiting either their X-points or the stochasticity resulting from their superposition \cite{feng_iaea08}.

The reversed field pinch (RFP) is the third toroidal configuration which could provide the basis for a fusion reactor \cite{rfp_generale,ortolani}. 
It presents some very attractive features, such as the weak requirements on toroidal field coil current and the possibility of reaching ignition with ohmic heating only, which could give rise to a simple, compact and relatively inexpensive reactor. However, until recently the RFP was considered to be flawed because of the presence of several resistive MHD modes in the saturated stage, which gave rise to an ergodization of the magnetic field over most of the plasma volume. These modes were considered to be intrinsic to the configuration, because they are responsible of the dynamo process, which allows the sustainment of the poloidal currents flowing in the plasma against resistive decay. 

This picture is now radically changing, thanks to the realization that the dynamo can be achieved with only one mode \cite{cp,finn}. This condition is named Single Helicity (SH) state, as opposed to the Multiple Helicity (MH) one. The theoretical prediction of the SH condition has been confirmed by experimental observation \cite{prl_qsh}. In the RFX-mod device Quasi Single Helicity (QSH) states, where one mode dominates over the others, have been observed to occur more and more frequently and for longer time duration as the plasma current is raised \cite{np}. This result is conditioned to a good control of the radial magnetic field at the edge, which in RFX-mod is obtained through a sophisticated system of 192 feedback-controlled saddle coils \cite{feedback,zanca09}.

In particular, interest in the RFP has been revamped by the discovery of the spontaneous transition at high plasma current to a new magnetic topology, dubbed Single Helical Axis (SHAx) state \cite{lorenzini_prl08,valisa,np}. In this state, which is a special instance of the more general QSH condition, the separatrix of the dominant mode magnetic island is expelled, while its X-point collapses onto the main magnetic axis of the discharge. The resulting configuration is characterized by having a helical magnetic axis, which is the former island O-point. In this condition order emerges from magnetic chaos in the core plasma and a transport barrier develops, so that the confinement properties of the configuration are enhanced \cite{np}. 

The SHAx observation, and the realization that it is going to be the standard condition in multi-MA RFP devices, have given new appeal to the RFP as an alternative configuration for controlled fusion. Some important issues have now to be faced, and in particular the problem of proper control of plasma-wall interaction, since up to now RFP devices have been conceived simply using the vacuum chamber wall as a limiter. The only exception to this are the STE-2 device, where a poloidal divertor has been tested \cite{STE2_poloidal}, and the TPE-2M device, where both toroidal and poloidal divertor configurations have been tried \cite{TPE2M_poloidal, TPE2M_toroidal}. In all these cases, no significant improvements in plasma performance were achieved.

In this paper we perform a detailed examination of the intrinsic edge magnetic topology in SHAx states obtained in the RFX-mod device \cite{sonato}. RFX-mod is the largest machine in the world operating in RFP configuration, with a maximum design plasma current of 2 MA. It has a major radius of 2 m and a minor radius of 0.459 m. The vacuum vessel has a circular cross section, and is internally covered by graphite tiles. Since the machine has no limiters, these tiles are intended to limit the plasma and protect the vessel. The presented results provide an answer to the still open question of why plasma-wall interaction in RFX-mod high current operation appears to be lower at shallow reversal. Furthermore, they allow to conclude that its intrinsic properties of the magnetic configuration could be exploited to construct a divertor, similar to the island divertor of stellarators. While further theoretical studies need to be done, and a practical implementation has to be demonstrated, we think that this is a conceptual breakthrough which reinforces the case of the RFP as a candidate reactor configuration and brings a useful contribution to magnetic fusion physics.

The paper is organized as follows: in Section \ref{mhd} the properties of the MHD spectrum measured in SHAx states which are relevant to our discussion are described; in Section \ref{concept} the topology of the edge plasma, and in particular the position and shape of the LCFS, are investigated by means of a field line tracing code; in Section \ref{pwi} the consequences of the edge magnetic topology on the plasma-wall interaction are studied; finally, in Section \ref{conclusions} conclusions are drawn.

\section{MHD mode spectrum in SHAx states}
\label{mhd}
The m=1 mode spectrum in SHAx states obtained in RFX-mod at the 1.5 MA plasma current level displays a dominant mode (the n=7 one) and weak amplitudes of the secondary modes \cite{pop_invito}. However, it important to notice that also m=0 modes are resonant in the RFP, and have their resonance radius located on the toroidal field reversal surface. This can be positioned more or less distant from the first wall, depending on the current imposed in the toroidal field coil. The distance from the wall is indirectly quantified by the reversal parameter $F=B_\phi(a)/\langle B_\phi\rangle$, where $\langle\ldots\rangle$ is an average over the poloidal cross section. 
A crucial observation made in RFX-mod is that shallow (slightly negative) F, which corresponds to a small distance of the reversal surface from the wall, turns out to be a better condition at high plasma current as far as plasma-wall inteaction is concerned than deeper (more negative) F. In fact, at plasma currents beyond 1 MA deep F discharges display sudden increases in density, probably induced by excessive wall heating which triggers the release of hydrogen from the graphite. These events in turn cause an increase of plasma resistance which makes it difficult to obtain a steady flat-top. Furthermore, in this condition the wall is rapidly loaded of hydrogen, so that the density control becomes impossible. This has led to the conclusion that shallow F operation, where this phenomenology is not observed, constitutes the optimal condition for high current operation. An explanation of this empirical finding will be given in the following. 

\begin{figure}
\centering
\includegraphics[width=0.6\columnwidth]{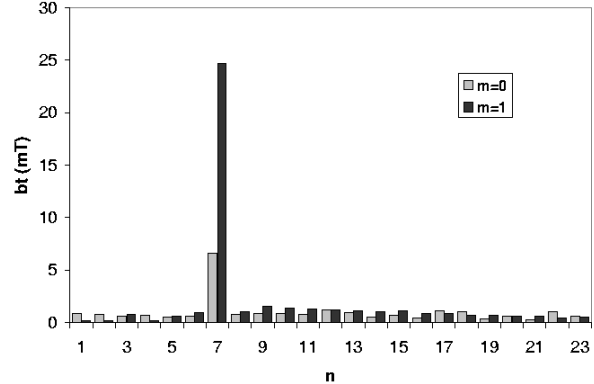}
\caption{Average spectra of m=0 and m=1 modes in 1.5 MA SHAx states. The mode amplitudes are evaluated on $B_\phi$ measurements performed outside the plasma.}
\label{spettri}
\end{figure}
\begin{figure}
\centering
\includegraphics[width=0.6\columnwidth]{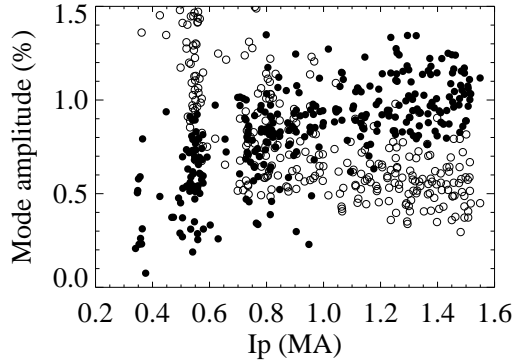}
\caption{Amplitude of the m=0/n=7 mode (full circles) and of the other m=0 modes up to n=15 (empty circles) in QSH conditions plotted as a function of plasma current. The mode amplitudes are normalized to the average poloidal field at the wall}
\label{scaling_modi_m0}
\end{figure}
\begin{figure}
\centering
\includegraphics[width=0.6\columnwidth]{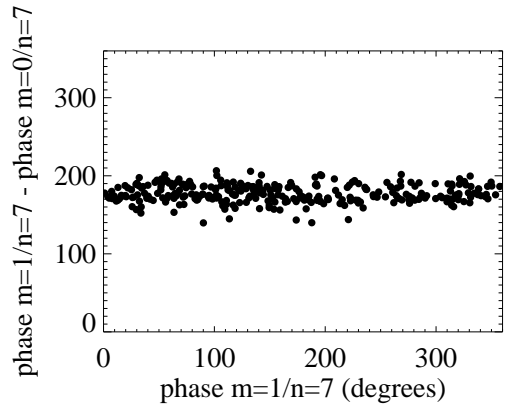}
\caption{Phase difference between the m=1/n=7 mode and the m=0/n=7 mode plotted as a function of the phase of the m=1/n=7 mode for SHAx states obtained in 1.5 MA discharges.}
\label{fasi}
\end{figure}

The m=0 mode spectrum, which features many low n modes in low current operation, is also strongly peaked on the n=7 one in SHAx states. This is shown in Fig. \ref{spettri}, where the average spectra of both m=0 and m=1 modes in SHAx states obtained at 1.5 MA plasma current are shown. The amplitudes are computed through a Fourier analysis of $B_\phi$ measurements obtained outside the plasma. 
The prevailing of the m=0/n=7 mode as current is increased is a direct consequence of the same behaviour observed for the m=1/n=7 mode \cite{np}, due to the toroidal coupling of the two modes. It is crucial to notice, as shown in Fig. \ref{scaling_modi_m0}, that the m=0 spectrum displays the same behaviour with increasing plasma current of the m=1 one, i.e. the dominant n=7 mode increases its relative amplitude while the secondary mode amplitude is reduced. 
Thus, it is possible to conclude that the RFP evolves, as plasma current is increased, towards a SHAx state characterized by the presence of a dominant m=1 mode and a dominant m=0 mode, both having the same n number. Furthermore, due to toroidal coupling, the two modes have a very clear phase relationship. 
This is  displayed in Fig. \ref{fasi}, where the phase difference between the m=0/n=7 mode and the m=1/n=7 mode in RFX-mod is shown as a function of the phase of the m=1/n=7 mode for a large set of discharges, in time intervals where a strong quasi-single helicity state is observed. It can be seen that the two modes have an almost constant phase difference equal to $\pi$.

The m=0 modes give rise to a chain of magnetic islands across the reversal surface, that is in the first wall proximity. In the past, in low current, multiple helicity conditions these islands actually acted as a transport barrier due to the very low degree of order of the surrounding plasma \cite{spizzo}. This situation has changed since the introduction of the active saddle coil system for the control of the radial magnetic field at the plasma boundary \cite{feedback}. This has lowered the tearing mode amplitude and created a more ordered situation, so that good conserved magnetic surfaces exist in the outer plasma region. Indeed, the m=0 islands have been recently found to be a source of flattening of the local temperature profile, due to their effect of short-circuiting different plasma radii \cite{vianello_eps08}.

\section{Topology of the edge region in SHAx RFP plasmas}
\label{concept}
In order to understand the magnetic topology of the edge region in SHAx states,  and in particular the position and shape of the LCFS, we have used a field-line tracing code named FLiT \cite{flit} to trace the magnetic topology in the plasma edge. FLiT uses the output of an algorithm for the reconstruction of the tearing mode eigenfunctions over the whole plasma volume based on Newcomb's equation, supplemented with edge magnetic measurements to account for the unknown derivative jumps in the resistive layers \cite{zanca_terranova}.

\begin{figure}
\centering
\includegraphics[width=0.6\columnwidth]{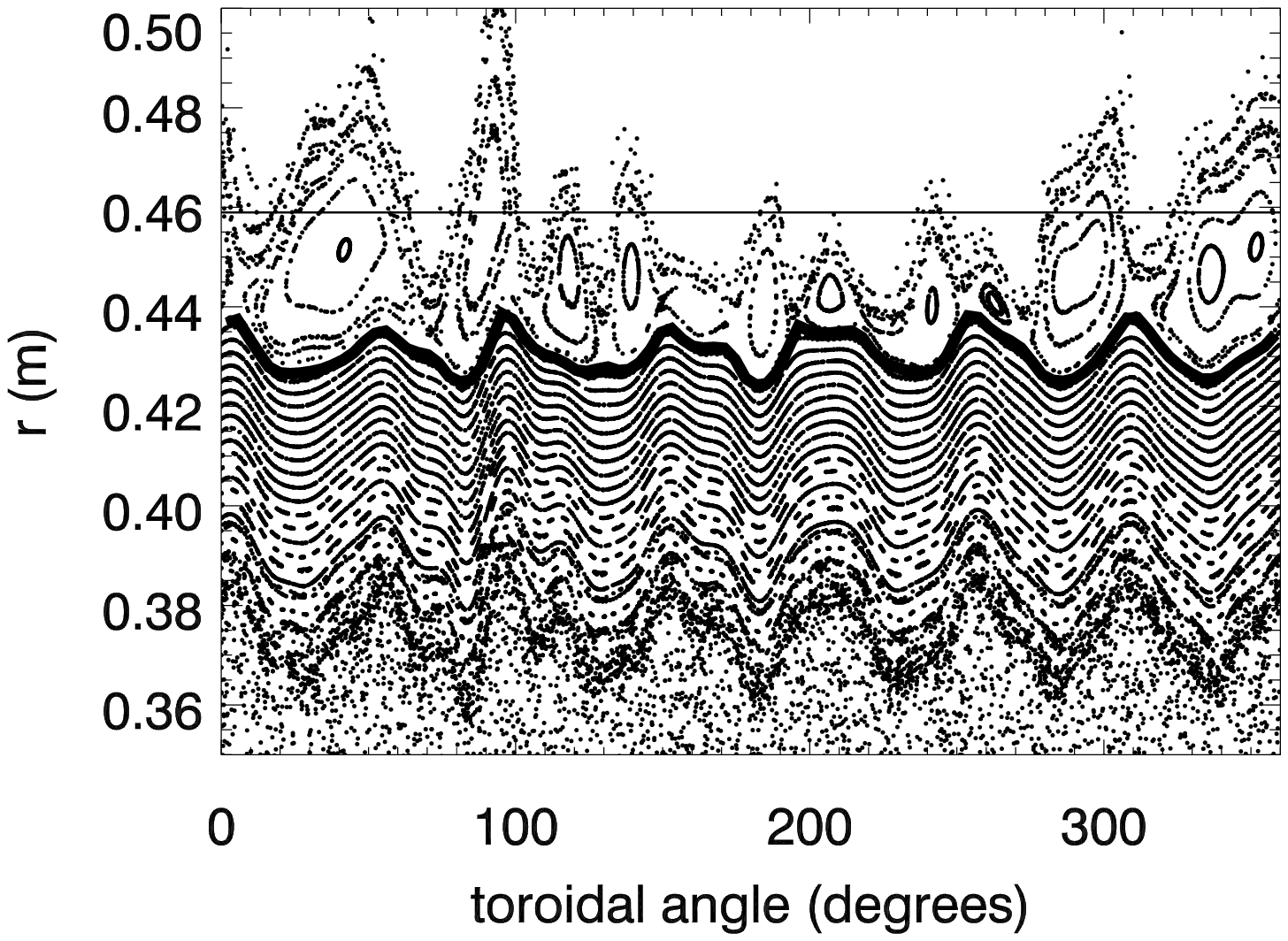}
\includegraphics[width=0.6\columnwidth]{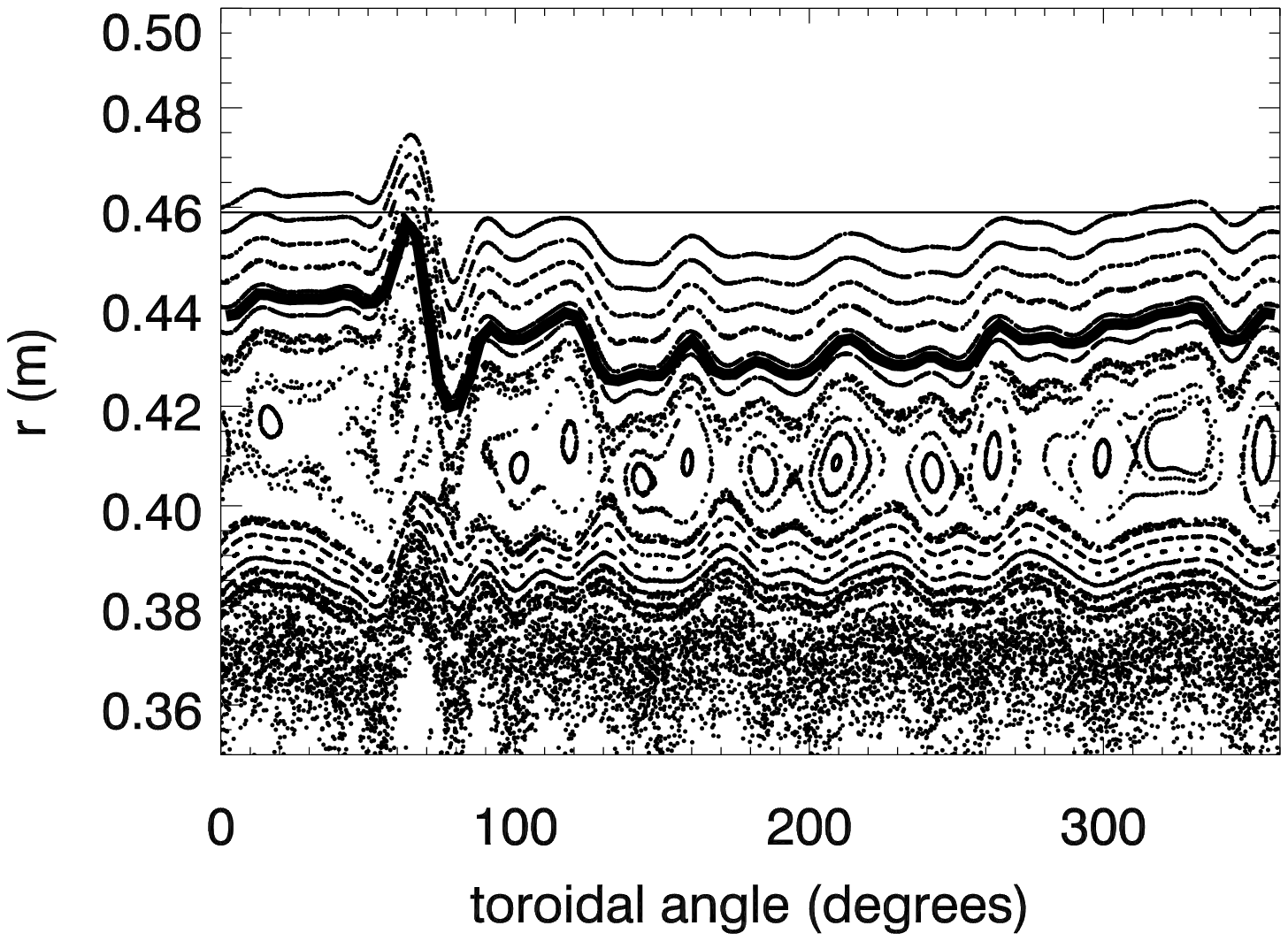}
\caption{Poincar\'e plot of the magnetic field lines on the outer equator for a SHAx state at 1.5 MA and shallow reversal (top) and for a similar condition at deep reversal (bottom). The thick line marks the position of the LCFS. while the horizontal line at $r=0.459$ m indicates the first wall position.}
\label{poincare}
\end{figure}

Fig. \ref{poincare} displays two Poincar\'e plots of the magnetic field lines in the r-$\phi$ plane on the outer equator (it is worth reminding the reader that in the outer region of RFPs the main magnetic field is almost poloidal). The first one is obtained during a 1.5 MA SHAx state at shallow reversal ($F = -0.017$), while the second depicts a similar condition obtained at deep reversal ($F=-0.181$). Thick lines are superposed on the plots, depicting the position of the LCFS, computed from FLiT outputs by looking at where, for each toroidal and poloidal position, the most internal open field line is found. 
The new and striking result is that in the SHAx condition obtained at shallow reversal the LCFS is well separated from the wall by the $m=0$ islands, and that their X-points act so as to form a divertor-like configuration. On the contrary, at deep reversal the LCFS is located beyond the m=0 island chain and a limiter-like condition is obtained. 

\begin{figure}
\centering
\includegraphics[width=0.6\columnwidth]{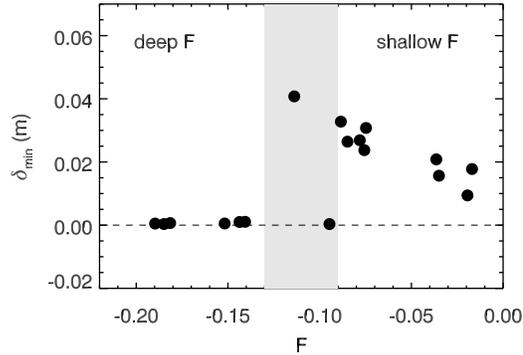}
\caption{Minimum distance of the LCFS from the first wall plotted as a function of the reversal parameter. The shaded region marks the F range where transition from a limiter-like geometry to a divertor-like one occurs.}
\label{deltamin_vs_f}
\end{figure}

The relationship between the occurrence of a divertor-like configuration and the reversal parameter value has been investigated more systematically by computing the minimum distance $\delta_{min}$ of the LCFS from the wall for a set of SHAx states obtained at different F values. 
Fig. \ref{deltamin_vs_f} displays a plot of $\delta_{min}$ as a function of F. It is clearly seen that as F goes from zero towards more negative values, that is from shallow reversal to deep reversal, the LCFS distance from the wall is increased, up to values of F between -0.10 and -0.13, where the m=0 islands do not intersect the wall any more and the limiter-like situation ($\delta_{min}=0$) is obtained. 
This is an explanation for the empirical evidence of a reduced plasma-wall interaction in shallow F discharges, which constitute the preferred mode of operation for RFX-mod at high current, as explained in section \ref{mhd}.

\section{Plasma-wall interaction}
\label{pwi}
\begin{figure}
\centering
\includegraphics[width=0.6\columnwidth]{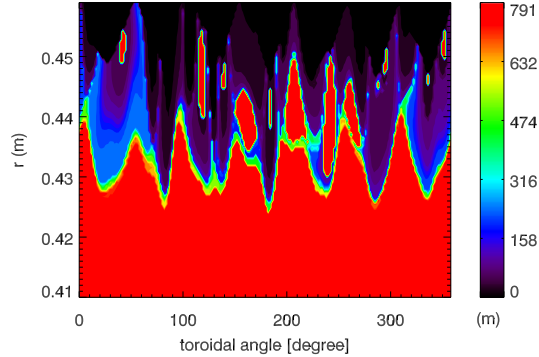}
\caption{Map of the correlation lengths on the r-$\phi$ plane located at $\theta=0$.}
\label{laminar}
\end{figure}

In order to better understand the structure of the Scrape-Off Layer (SOL) formed beyond the LCFS in the shallow F case shown in the top frame of Fig. \ref{poincare}, a colour scale plot of the connection lengths of field lines passing through a grid of points in the r-$\phi$ plane has been constructed. This has been done integrating the field line equation from the starting point, both forwards and backwards, until the first wall is reached. The connection length is then defined as the sum of the two lengths covered in the two directions.
The result of the procedure is shown in Fig. \ref{laminar}, for the plane at $\theta=0^\circ$. The maximum value of the connection length, for which the red colour has been used, also marks the closed field lines. The confined plasma, enclosed by the LCFS modulated by the n=7 pattern, can be clearly identified. Furthermore, other red regions, corresponding to m=0 islands that do not touch the first wall, can be observed. These could be associated to the high radiation poloidal rings which have been proposed to explain the RFP density limit \cite{anelli_radiativi}. Beyond the LCFS a SOL is created. 
In particular, in the last cm near the first wall relatively short connection lengths are found, with only occasional regions of longer connection lengths reaching the wall.

\begin{figure}
\centering
\includegraphics[width=0.5\columnwidth]{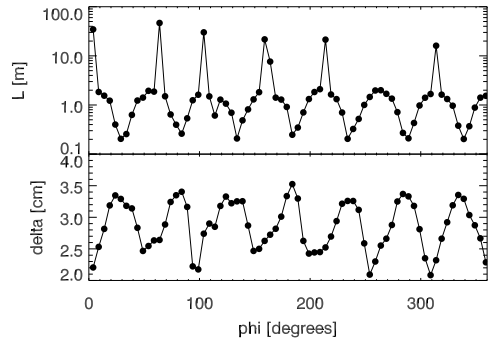}
\caption{Connection lengths of field lines originating from different points of the first wall ($r=0.459$ m) on the outboard equatorial plane ($\theta=0$) and distance between the LCFS and the wall on the same plane, both plotted as a function of the toroidal angle.}
\label{Lconn_delta}
\end{figure}

The connection length of the field lines touching different points of the first wall is closely related to the local distance between the LCFS and the wall. This is shown in Fig. \ref{Lconn_delta}, where the connection length of points of the first wall located on the outer equator is shown as a function of the toroidal angle, for the same condition of the top frame of Fig. \ref{poincare}. 
It is possible to observe that the connection length displays an $n=7$ periodic structure, with rounded maxima which are anyway lower than the length of one poloidal turn ($\sim 3$ m). Superimposed to this, limited regions of much larger connection lengths are found, indicating field lines which manage to perform many poloidal turns before touching again the wall. The origin of these long field lines are positioned on the maxima of the $n=7$ pattern.
In the same figure is also plotted the distance between the LCFS and the first wall, again plotted as a function of the toroidal angle for points along the outer equator. The $n=7$ periodicity can also be clearly seen. 
The comparison of the two curves allows to conclude that in the positions where long connection lengths exist, the LCFS-wall distance is shorter, while in regions of short connection length the LCFS-wall distance is larger. The consequence that can be drawn from this fact, when thinking to the plasma-wall interaction, is that in the first situation a stronger interaction is expected, both because of parallel flows, since a longer flux tube collects a larger amount of energy, and because of perpendicular fluxes, which originate from the LCFS which is less distant.

\begin{figure}
\includegraphics[width=\columnwidth]{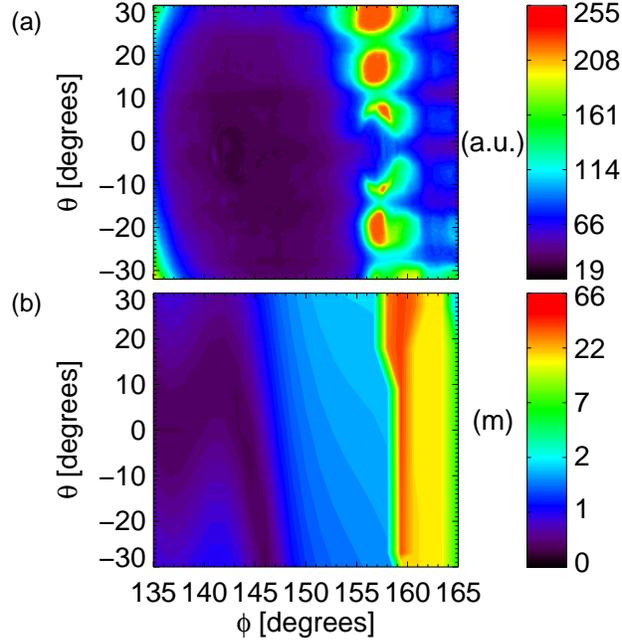}
\caption{Pattern of $H_\alpha$ emission recorded by a fast camera and mapped onto a portion of the poloidal-toroidal plane (a). Map of the connection lengths of the field lines originating from the first wall for the same discharge and time instant (b).}
\label{camera}
\end{figure}

The hypothesis that regions of the first wall with larger connection length have a stronger plasma-wall interaction has been validated using data from a fast CCD camera looking at $H_\alpha$ line emission. This radiation originates from neutral hydrogen atoms coming out from the wall. Since in RFX-mod the density profile displays a strong gradient in the first few cm of the plasma, it is possible to associate the emission to the heat load of the nearby portion of the first wall, under the assumption that a stronger plasma-wall interaction heats locally the wall and causes a stronger hydrogen release. The camera used in RFX-mod was operated with a frame rate of 10,000 frames per second, and a shutter time of 1/10,000 s.
Fig. \ref{camera}a shows the emission pattern detected by the camera in a 1.5 MA discharge, during a SHAx phase. The actual image has been remapped to a regular grid in the the toroidal and poloidal angles. The figure displays an almost vertical emission pattern, with a discretization which corresponds to the tiles which compose the first wall. This discretization is due to the tile shape, which causes an increased interaction in the central part of the tile. A darker region in the middle of the bright pattern (around $\theta=0^\circ$) is due to the presence of a port, as it is the dark oval region more on the left (around $\phi=142^\circ$). 
In Fig. \ref{camera}b the connection lengths for the same region of the first wall, in the same discharge and at the same time instant are plotted. 
It is clearly seen that the connection length pattern is similar to the emission pattern of the camera. The two patterns are displaced one with respect to the other by a few degrees in the toroidal direction. This discrepancy appears in all cases, and is attributed to a systematic error. 

\begin{figure}
\centering
\includegraphics[width=0.5 \columnwidth]{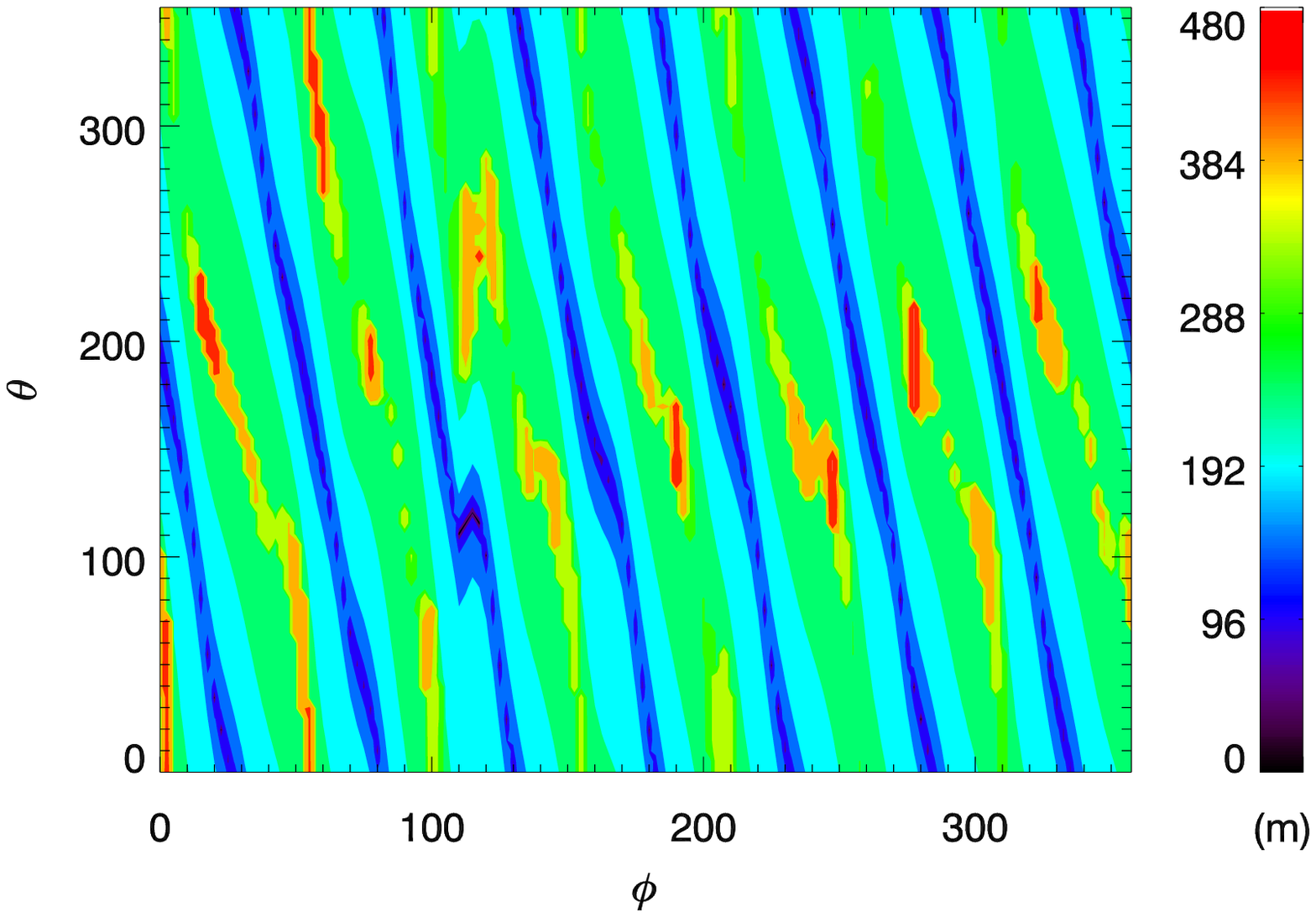}
\includegraphics[width=0.5 \columnwidth]{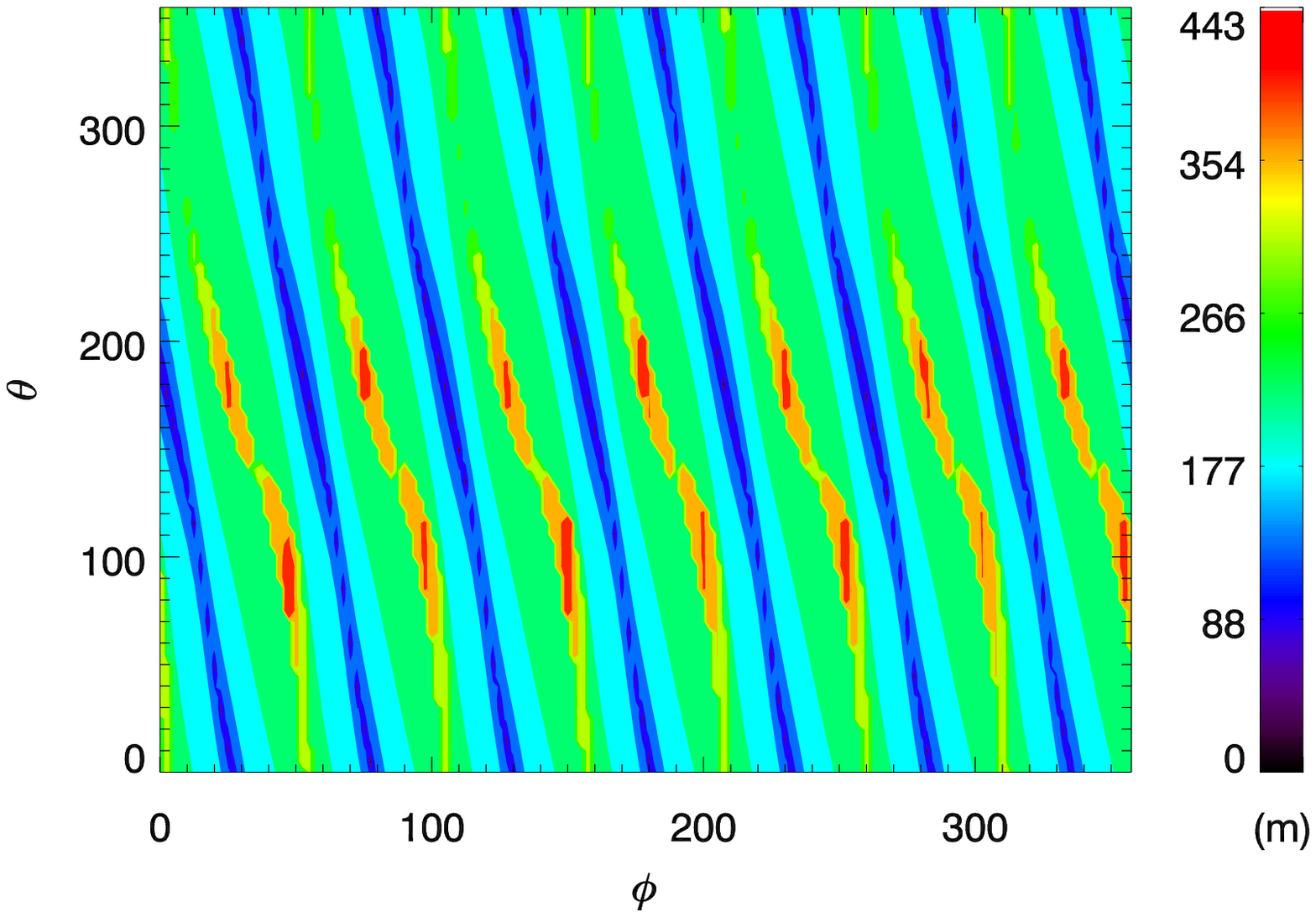}
\caption{Map of the connection lengths of magnetic field lines touching the first wall for a typical 1.5 MA SHAx state (top) and same map obtained using only the amplitudes of the dominant m=1/n=7 and m=0/n=7 modes, indicating the ideal situation for a pure single helicity condition (bottom).}
\label{wallmap}
\end{figure}

Having established that the magnitude of the plasma-wall interaction in different regions of the first wall can, to a first approximation, be quantified by looking at the connection length of field lines leaving them, it is possible to build a map of this interaction by following field lines starting from a grid covering the first wall. The top frame of Fig. \ref{wallmap} shows such a map, with the connection lengths represented as a colour scale. It can be seen that regions of higher connection lenghts concentrate in the $90^\circ$-$180^\circ$ band, that is in the region comprised between the torus top and the inner equator. 
This is a non-trivial consequence of the phase relationship between the m=0 and m=1 modes depicted in Fig. \ref{fasi}. The localization is not perfect, due to the polluting effects of the secondary modes.
 
In order to understand what is the tendency for higher current plasmas, where the secondary modes will be lower according to present scalings, we have performed the same calculation including only the m=1/n=7 and m=0/n=7 modes in the FLiT input, that is simulating a pure single helicity condition. The result is shown in the bottom frame of Fig. \ref{wallmap}. It can be seen that the long connection length region is now limited to two rows of inclined regions having an n=7 periodicity, the first located at $\theta\simeq 90^\circ$ and the second at $\theta\simeq 180^\circ$. 

\section{Conclusions}
\label{conclusions}
The results described in this paper give a new vision of the plasma boundary in the high performance SHAx states recently discovered in high current RFP plasmas produced on the RFX-mod device. Due to the toroidal coupling between the dominant $m=1/n=7$ mode, responsible for the achievement of a helical equilibrium, and its $m=0$ counterpart, a set of $m=0$ magnetic islands with a dominant $n=7$ periodicity is formed on the reversal surface. 
Provided that the absolute value of the reversal parameter is small enough, that is the machine is operated with a shallow reversal, these islands intercept the first wall. Due to the regularity of the SHAx condition, where no strong localized distortion of the plasma column due to mode locking \cite{locking} is found, the outcome is that the LCFS is not any more touching the first wall. On the contrary, the X-points formed between the $m=0$ islands act as the X-point of a tokamak divertor. This condition is reminiscent of the island divertor concept which is being explored as a means of controlling plasma-wall interaction in stellarators. 

Indeed, the plasma-wall interaction in these discharges is strongly affected by the edge region topology. The LCFS shape is modulated by an $n=7$ displacement, and the positions where it is less distant from the first wall coincide with locations where field lines touching the first wall have a very long connection length. Thus, from the point of view of both parallel and perpendicular transport these are the regions of stronger interaction. This has been confirmed experimentally using data from a fast camera looking at $H_\alpha$ emission.

Following these findings, one is led to infer that the special features of the plasma boundary in high current SHAx RFP plasmas could be exploited for building a divertor, similar in concept to the island divertor used in some stellarators. This could be achieved by locating divertor plates with appropriate pumping in the regions of strong interaction, which have been found to become more and more regular as the amplitude of the secondary modes is reduced, which is the trend experimentally observed as plasma current is increased. Such an approach to plasma-wall interaction would require that the dominant $m=0$ mode is stationary in time, but this appears to be a requirement that can be fulfilled, given the strong phase relationshop with the dominant $m=1$ mode, whose phase has been shown to be indeed feedback-controllable.
While a great deal of work is still required, especially concerning the details of transport from the LCFS to the different regions of the first wall, it is our hope that this work can be used as a starting point for the solution of one of the big open issues in the path of the RFP towards reactor feasibility.

\ack
This work, supported by the European Communities under the contract of Association between EURATOM/ENEA, was carried out within the framework of the European Fusion Development Agreement.


\section*{References}
\begin{thebibliography}{10}
\bibitem{stangeby} Stangeby P. C. 2000 \textit{The plasma boundary of magnetic fusion devices} (IOP Publishing, Bristol)
\bibitem{wesson} Wesson, J. 2004 \textit{Tokamaks, 3rd Edition} (Oxford Univesity Press, Oxford)
\bibitem{review_divertor_stellarator} K{\"o}nig R. \etal 2002 \PPCF \textbf{44} 2365
\bibitem{feng_iaea08} Feng Y. \etal, in Proc. of the 22nd IAEA Fusion Energy Conference, Geneva, 2008, TH/P4.4 (to be published).
\bibitem{rfp_generale} Bodin H. A. B. and Newton A. A. 1980 \NF \textbf{20} 1255
\bibitem{ortolani} Ortolani S. and Schnack D. D. 1993 \textit{Magnetohydrodynamics of Plasma Relaxation} (World Scientific, Singapore)
\bibitem{cp} Cappello S. and Paccagnella R. 1992 \PFB \textbf{4} 611
\bibitem{finn} Finn J. M., Nebel R. and Bathke C. 1992 \PFB \textbf{4} 1262
\bibitem{prl_qsh} D. F. Escande, Martin P., Ortolani S. \etal 2000 \PRL \textbf{85} 1662
\bibitem{np} Lorenzini R., Martines E., Piovesan P. \etal 2009 \textit{Nature Phys.} \textbf{5} 570
\bibitem{feedback} Zanca P., Marrelli L., Manduchi G. and Marchiori G. 2007 \NF \textbf{47} 1425
\bibitem{zanca09} Zanca P. 2009 \PPCF \textbf{51} 015006
\bibitem{lorenzini_prl08} Lorenzini R., Terranova D., Alfier A. \etal 2008 \PRL \textbf{101} 025005
\bibitem{valisa} Valisa M., Bolzonella T., Buratti P. \etal 2008 \PPCF \textbf{50} 124031 
\bibitem{STE2_poloidal} Iida M., Miyagi T., Ishijima D. \etal 1994 \PPCF \textbf{36} 153
\bibitem{TPE2M_poloidal} Hattori K., Hayase K. and Sato Y. 1994 \textit{J. Phys. Soc. Jpn} \textbf{63} 2177
\bibitem{TPE2M_toroidal} Hattori K., Hayase K. and Sato Y. 1995 \NF \textbf{35} 981 
\bibitem{sonato} Sonato P., Chitarin G., Zaccaria P. \etal 2003 \textit{Fusion Eng. Des.} \textbf{66} 161
\bibitem{pop_invito} Lorenzini R., Agostini M., Alfier A. \etal 2009 \PP \textbf{16} 056109
\bibitem{spizzo} Spizzo G., Cappello S., Cravotta A. \etal 2006 \PRL \textbf{96} 025001
\bibitem{vianello_eps08} Vianello N., Martines E., Agostini M. \etal 2009 \NF \textbf{49} 045008
\bibitem{flit} Innocente P., Alfier A., Carraro L. \etal, \NF 2007 \textbf{47} 1092
\bibitem{zanca_terranova} Zanca P. and Terranova D. 2004 \PPCF \textbf{46} 1115
\bibitem{anelli_radiativi} Puiatti M. E., Scarin P, Spizzo G. \etal 2009 \PP \textbf{16} 012505
\bibitem{locking} Zanca P., Martines E., Bolzonella T. \etal 2001 \PP \textbf{8} 516
\end{thebibliography}
\end{document}